\documentclass[%
 reprint,
 amsmath,amssymb,
 aps,
floatfix
]{revtex4-1}

\usepackage{graphicx}
\usepackage{dcolumn}
\usepackage{bm}

\begin{document}

\preprint{}

\title{Finite response time in stripe formation by bacteria with density-suppressed motility}

\author{Xingyu Zhang}
\altaffiliation[Present address: ]{Department of Biology, Friedrich-Alexander-Universit\"at Erlangen-N\"urnberg, Erlangen, Germany. / Max-Planck-Zentrum f\"ur Physik und Medizin, Erlangen, Germany.}
\author{Namiko Mitarai}%
 \email{mitarai@nbi.dk}
\affiliation{%
The Niels Bohr Institute, University of Copenhagen,
Blegdamsvej 17, 2100-DK, Denmark.
}%

\date{\today}

\begin{abstract}
Genetically engineered bacteria to increase the tumbling frequency
of the run-and-tumble motion for the higher local bacterial density form
visible stripe pattern composed of successive high and low density regions on an agar plate. We propose a model that includes a simplified regulatory dynamics of the tumbling frequency in individual cells to clarify the role of finite response time. We show that the time-delay due to the response dynamics results in the instability in a homogeneous steady state allowing a pattern formation. For further understanding, we propose a simplified two-state model that allows us to describe the response time dependence of the instability analytically. We show that the instability occurs at long wave length as long as the response time is comparable with the tumbling timescale and the non-linearity of the response function to the change of the density is high enough. The minimum system size to see the instability grows with the response time $\tau$, proportional to $\sqrt{\tau}$ in the large delay limit. 
\end{abstract}

\pacs{Valid PACS appear here}
\maketitle


\section{\label{sec:intro}Introduction}
Pattern formation in actively moving and growing living systems have been attracting interest in the physics community in the last decades \cite{fujikawa1989fractal,ben2000cooperative,shapiro1998thinking}. 
Especially, bacterial systems provide unique opportunities to perform various well-controlled experiments thanks to their fast dynamics and growth as well as the possibility of engineering the molecular network to control their action. 
One of the interesting works has been done on the pattern formation by bacterial motion and growth, where a theoretical analysis of growing bacteria population with density dependent motility \cite{cates2010arrested} predicted various arrested patterns, including concentric ring of high and low bacterial density regions. This  was experimentally observed by Liu {\it et al.} \cite{liu2011sequential}, where {\it Escherichia coli} was engineered to combine the quorum sensing circuit  of {\it Vibrio fisherei} \cite{waters2005quorum} with {\it E. coli}'s chemotaxis regulatory network \cite{berg1972chemotaxis,berg2008coli}. 

{\it E. coli} swims by the run-and-tumble motion, i.e., a bacterium "runs" roughly straight at a constant speed (about 10-20 $\mu$m/sec\cite{alon1998response}) between the "tumbling" where the direction of the swimming is shuffled randomly. In a wild type {\it E. coli}, the tumbling frequency decreases (increases) upon the increase (decrease) of attractant (repellant), but when a bacterium is left in a constant ligand environment, the tumbling frequency shows adaptation to go back to the original tumbling frequency. 
This change of the tumbling frequency and adaptation enable {\it E. coli} to move towards higher (lower) concentration of the attractant (repellant) \cite{barkai1997robustness,alon1999robustness}. The regulation is triggered by the binding of the ligand to the receptor. The binding affects the phosphorylation of the protein CheY, which interacts with the flagella motor to change the tumbling frequency. Previous research has shown that the change of CheY concentration keeps the tumbling duration to be constant and short ($\sim$ 0.2 second) while changing the running duration from sub-second to $\sim$10 seconds \cite{alon1998response}. Phosphorylated CheY is de-phosphorylated by protein CheZ in its free form \cite{bren1996signal}, which is required for rapid response of the cells. 

In the engineered bacteria \cite{liu2011sequential}, CheZ protein level was regulated to respond to the local cell density. More precisely, the cells were modified to produce and recognize the quorum sensing molecule of  {\it Vibrio fisherei} \cite{waters2005quorum}. The accumulation of the quorum sensing molecules in the environment due to high local density of the bacteria activates the production of a repressor protein CI,  which represses the production of the protein CheZ that is negatively correlated with the tumbling frequency. Thus the engineered bacteria tumble more frequently in the dense region, reducing the diffusion constant of the cells. When such bacteria is grown on a agar plate, the sequential establishment of stripe pattern of the high- and low-cell density regions were observed \cite{liu2011sequential}.  

This system has been previously modeled by using a reaction-diffusion type equations that describes the bacteria motility through the diffusion and the growth term is added to it \cite{liu2011sequential,fu2012stripe}. One of the important assumptions in the model is that the bacterial density $\rho$ obeys the following type of the equations:  
\begin{equation}
\frac{\partial \rho}{\partial t}=\nabla^2 \mu(h)\rho +g(\rho)
\label{eq:modelorig}
\end{equation}
where $\mu(h)$ is the quorum sensing molecule density ($h$) dependent bacterial motility, and $g(\rho)$ is the bacterial growth rate.
The equation (\ref{eq:modelorig}) assumes that the diffusion flux of the bacteria $J$ is given by $J=-\nabla \mu(h)\rho$, with the quotum-sensing molecule dependent motility being inside the gradient. If in contrast one assumes $J=-\mu(h)\nabla \rho$, the uniform density is the only solution for no flux steady state $J=0$, thus the model would not support the pattern. Clearly, in this case, the uniform density state is the only no-flux steady state without growth ($g(\rho)=0$).

It is not trivial when $J$ should take the form  $J=-\nabla \mu(h)\rho$. Indeed, the previous works \cite{schnitzer1993theory,tailleur2008statistical} derived the following argument:
For simplicity, let's consider a one dimensional run-and-tumble model, where a bacterium run $+$ or $-$ direction with a {\it constant} speed $v$, and it changes its direction with the frequency $\alpha(h)$ (i.e., the tumbling duration is ignored), where $h$ is the signaling molecule density that the tumbling frequency responds to.
Denoting the population that moves to the $+$ ($-$) direction as $p_+$ ($p_-$), we have 
\begin{eqnarray}
\frac{\partial p_+}{\partial t}&=&-v\frac{\partial p_+}{\partial x}-\alpha(h)p_++\alpha(h)p_-, \label{eq:p+}\\
\frac{\partial p_-}{\partial t}&=&v\frac{\partial p_-}{\partial x}+\alpha(h)p_+-\alpha(h)p_-.
\end{eqnarray}
Then the total density $\rho=p_++p_-$ obeys
\begin{eqnarray}
\frac{\partial \rho}{\partial t}&=&-\frac{\partial J}{\partial x},\label{eq:rho}\\
\frac{\partial J}{\partial t}&=&=-v^2\frac{\partial \rho}{\partial x}-2\alpha(h)J,\label{eq:J}
\end{eqnarray}
with $J\equiv v(p_+-p_-)$. 
When we assume the relaxation of flux is fast hence $\partial J/\partial t=0$, we get $J=-v^2/(2\alpha(h)) \nabla \rho$.  
Clearly, the no-flux steady state $J=0$ supports only the uniform solution of a constant $\rho$, even though the effective diffusion constant $D=v^2/(2\alpha(h))$ depends on a parameter $h$ that may vary over space. 

Of course, the run-and-tumbling model described above is a simplification of the reality, and the macroscopic model of type eq.~(\ref{eq:modelorig}) may be derived from a more realistic microscopic model. For example, it is well known that, if the running velocity $v$ is depends on $h$, the above calculation will be modified to allow the non-uniform steady state \cite{schnitzer1993theory,tailleur2008statistical}.  Another way is to take into account the finite duration of tumbling, which makes $h$ dependent tumbling frequency $\alpha$ to produce non-uniform steady states \cite{AgnereCuratolo2017}.

One of the significant simplification in the described model is to assume that the tumbling frequency would change immediately upon the change of the quorum-sensing molecules \cite{AndreasLund2015}. In the engineered bacteria, quorum-sensing molecule indirectly regulate the transcription of {\it cheZ} gene. The time scale for the change of the local density of the quorum-sensing molecule reflected in the CheZ protein level is determined by the lifetime of CheZ protein and intermediate regulator proteins CI. When the protein is stable, as often assumed for the repressor CI \cite{aurell2002stability} unless there is a DNA damage that activates the CI degradation \cite{ptashne2004genetic}, this response time scale can be as slow as the cell division time scale. This is in contrast to the wild type cell's chemotaxis regulation, where the signal transfer is fast because it is done by changing the protein activity by phospholyration and other modification, and not by changing the protein levels \cite{barkai1997robustness,berg2008coli}. For chemotaxis, 
many models have been proposed and analyzed under externally imposed chemical gradient \cite{oosawa1977behavior,lapidus1980pseudochemotaxis}, and even the self-aggregation via combination of quorum-sensing-like chemical and the chemotaxis has been studied \cite{inoue2008conditions}. However, the response time in these situation is rather fast, and the adaptation of the tumbling frequency is the important part of those modeling.  

Recently, a more detailed model of the experimental system \cite{liu2011sequential} has been analyzed by Xue {\it et al.} \cite{xue2018role}. 
After proposing model equations that take into account intracellular signaling including CheZ production and dilution by growth for each cell,  
corresponding partial differential equations were derived by assuming the separation of time scales of intracellular signaling processes, run-and-tumble motions, and growth-induced pattern formation. With using realistic parameter values, the model successfully reproduced the experimentally observed patterns. Interestingly, the authors analyzed the effect of CheZ turnover timescale in the pattern, and reported that the stripe would not form when the time scale of CheZ degradation is 10-fold slower than the dilution by cell growth, while 10-fold faster response of CheZ results in sharper patterns. 
This work clearly demonstrates the importance of the response delay, but the relative complexity of the model makes it difficult to obtain deeper understanding, such as the functional form of the delay-time dependence of the instability threshold. 

Motivated by these observations, we propose a simple run-and-tumble model with density dependent tumbling frequency with time delay in the response.
For simplicity, we set the growth term of the bacterium to be zero for most of the analysis, and focus on the instability of the uniform solution, which is necessary for the stripe formation. 
This model considers the dynamics of the internal state of a cell, and the internal state variable controls the tumbling frequency. 
We show that the model can support nonuniform density distribution as long as the time delay is finite, and we analyze the role of the time delay in supporting a non-uniform density profile. 
We then propose a simplified two-state model, where the bacteria can take only high or low tumbling frequency, but the ratio changes depending on the density with a time scale $\tau$. The model allows us to analytically determine the condition for the instability of the uniform solution, revealing that non-zero time delay and strong enough non-linearity can make uniform solution unstable within a certain range of density, but the system size required for the instability increases with the time delay. We study how this reflects in the stripe formation with growth by adding the growth term in the model, and show that the coupling between the protein dilution rate and the bacterial growth rate ensures that the pattern can be formed as long as the non-linearity of the response is strong enough.    

\section{\label{sec:fullmodel} Run-and-Tumble model with the regulator protein}
\subsection{Microscopic model}
First, we consider a microscopic model, where individual cell is moving either left or right at a speed $v$, and changes the direction with the tumbling frequency $\alpha(c_i)$. Here, $c_i$ is the concentration of the tumbling-frequency controlling protein in the $i$-th cell. We assume that the tumbling frequency is an increasing function of $c_i$, with
\begin{equation}
\alpha(c)=\alpha_l+\frac{\alpha_h-\alpha_l}{1+(c/c_0)^{-m}},
\label{eq:alphac}
\end{equation}
where $\alpha_l$ and $\alpha_h$ are the minimum and the maximum tumbling frequency, respectively, $c_0$ characterizes the threshold concentration of the regulatory protein for a cell to be in high tumbling mode, and $m>0$ characterizes the sharpness of the response. 
\begin{figure}[t]
\includegraphics[width=0.5\textwidth]{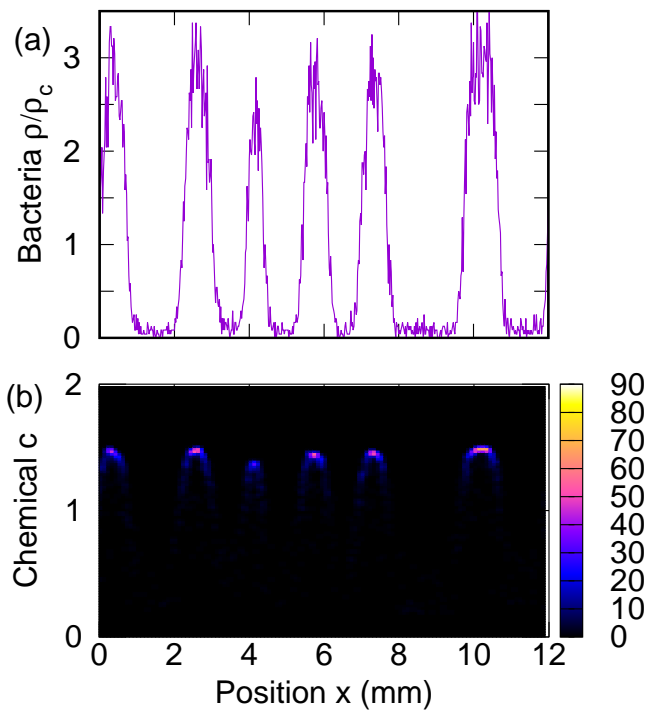}
\caption{A snapshot of the simulated microscopic model. (a) The bacterial density distribution over space. (b)The horizontal axis shows the position, and the vertical axis shows the concentration of the regulator $c$, and the color represents the probability density.}
\label{MicroscopicModelFig}
\end{figure}

The protein $c_i$ is transcriptionally regulated. The regulation is through the quorum-sensing system in reality, but here we make a simplification that the active regulator level is proportional to $\rho(t,x_i)$, where $x_i$ is the position of the bacterium $i$. 
Under these assumptions, we model the activation of $c_i$ production by the bacterial density as 
\begin{equation}
\frac{d c_i}{dt}=\frac{k_1}{1+(\rho(t,x)/\rho_c)^{-n}}-\frac{c_i}{\tau}.
\label{eq:c}
\end{equation}
Here, $k_1$ is the maximum production rate of $c_i$, and $n>0$ describes the cooperativity of the regulation, with the threshold density $\rho_c$ for activation. The degradation time scale $\tau$ determines the timescale of the response to the change of the bacterial density. Combined with eq.~(\ref{eq:alphac}), the bacteria in high density region will tumble more often, resulting in the smaller motility. 

We have simulated this one-dimensional model. In the following, we set the running velocity
$v=15\mu$m/s, the tumbling frequencies $\alpha_h=11.25$/s
and $\alpha_l=0.25/s$. This gives the effective diffusion constant $D$ to be within the range from 10$\mu$m$^2$/s to 450$\mu$m$^2$/s. 
The response time scale $\tau$ was set to $400s$, which is reasonably long compare to the time scale tumbling. The non-linearlity of the responses are set by choosing $m=10$ and $n=1$. Threshold protein level $c_0$ is considered as a concentration unit. We set $k_1=2c_0\tau$, which allows the maximum protein level to reach twice of the regulation threshold $c_0$. 
We discretized the space into meshes of the size $\Delta x=20\mu$m,  and the local density $\rho (x)$ was calculated by counting all the cells in the mesh and divided by mesh size. The density threshold $\rho_{c}$ was set to $1.2/\mu$m. The model was simulated by with a constant time step $\Delta t=0.005$s. 

Figure~\ref{MicroscopicModelFig} shows a snapshot of the simulation 
of $1.2\times 10^4$ cells in the system size $L=1.2\times 10^4\mu$m under the periodic boundary condition. At time zero, the cells are distributed uniformly in space with the protein level uniformly distributed between $0$ and $c_{max}$.  
The snapshot was taken after $10^5$s $\approx$ 28 h.
The figure clearly shows the development of locally high density regions and low density regions (Fig.~\ref{MicroscopicModelFig}a). 
Figure~\ref{MicroscopicModelFig}b shows that the cells in the high density region has high level of the protein $c_i$, resulting in the high tumbling frequency. This clearly depicts that with the finite response time introduced by the regulation (\ref{eq:c}) destabilizes the uniform density state.

\subsection{Continuum description of the model} 
Since the microscopic model is rather time consuming to simulate, we simulate the behavior in a wider range of parameters by using a continuum version of the model. We consider the the master equation of the distribution function $f(t,x,c)$, where $f(t,x,c)dxdc$ is the probability for bacteria to have regulator concentration between $c$ and $c+dc$ and to be at location between $x$ and $x+dx$. It is normalized so that $N=\int_{-\infty}^\infty \int_0^\infty f(t,x,c)dxdc$ gives the total number of the bacteria cells in the system. We further consider the noise in the protein concentration $c$ by interpreting eq.~(\ref{eq:c}) as a rate equation, in order to make $f(t,x,c)$ smooth in the $c$-space. Using the method parallel to those described in eqs.~(\ref{eq:p+})-(\ref{eq:J}) and performing the system size expansion for the noise in $c$ \cite{van1992stochastic}, we obtain
\begin{eqnarray}
\rho(t,x)&\equiv&\int_0^{\infty}f(t,x,c)dc, \label{rhovsf}\\
\frac{\partial f}{\partial t}&=&-\frac{\partial J_x}{\partial x}
-\frac{\partial}{\partial c}\left[f\left(\frac{k_1}{1+(\rho/\rho_c)^{-n}}-\frac{c}{\tau}\right)\right]\nonumber\\
&&+\frac{1}{2\Omega}\frac{\partial^2}{\partial c^2}\left[f\left(\frac{k_1}{1+(\rho/\rho_c)^{-n}}+\frac{c}{\tau}\right)\right], \label{rhoc}\\
\frac{\partial J_x}{\partial t}&=&-v^2\frac{\partial f}{\partial x}-2\alpha(c)J_x\nonumber\\
&&-\frac{\partial}{\partial c}\left[J_x\left(\frac{k_1}{1+(\rho/\rho_c)^{-n}}-\frac{c}{\tau}\right)\right]\nonumber\\
&&+\frac{1}{2\Omega}\frac{\partial^2}{\partial c^2}\left[J_x\left(\frac{k_1}{1+(\rho/\rho_c)^{-n}}+\frac{c}{\tau}\right)\right]. \label{jxc}
\end{eqnarray}
\begin{figure}[h]
\includegraphics[width=0.4\textwidth]{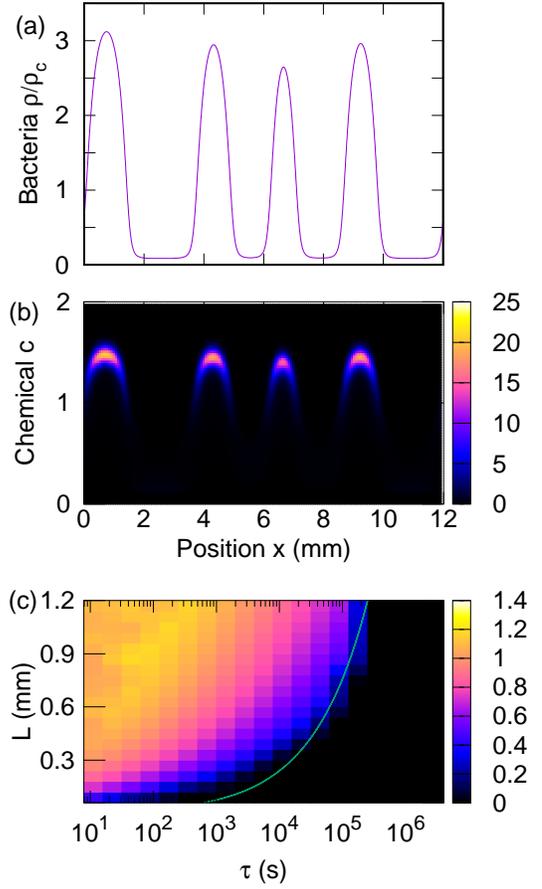}
\caption{Snapshot of the continuum model with the initial condition of uniform density $\rho_0=1$ with small randomness by adding uniformly distributed number in the range $(0, 6.25\times 10^{-8}\rho_0)$ to the initial density. (a) The snapshot of bacterial density over space at time $t_s=30000s$. (b)The horizontal axis shows the position, and vertical axis shows the concentration of the regulator $c$, and the color represents the probability density $f(t_s,x,c)$.  
(c) The phase diagram of the instability as a function of the time delay $\tau$ and the system size $L$. The system is simulated with the initial condition as in (a), and the instability was detected by calculating the standard deviation $\sigma$ of the density profile $\rho_f(x)$ after long time from the uniform solution, where $\sigma^2=\frac{1}{L}\int_0^L (\rho(x)-\rho_0)^2 dx$.
The solid line is proportional to $\sqrt{\tau}$, which is shown as a guide for the eye.}
\label{ContinuumModelFig}
\end{figure}

This equation have a uniform steady state solution for a given average density $\rho_0$, which is given by  
\begin{eqnarray}
f_{ss}(c)&=& A e^{-bc}(a+c)^{2ab-1},\\
a&=&\frac{c_{max}}{c_0}\frac{1}{1+(\rho_{0}/\rho_c)^{-n}},\\
b&=&2\Omega c_0.
\end{eqnarray}
Here, $A$ is a normalization constant, defined by  
$\int_0^\infty f_{ss}(c) dc=\rho_0$ (See eq.~(\ref{rhovsf})). 
We analyze the stability of this uniform solution by numerically simulating the model with an initial condition slightly perturbed from the uniform solution by small noise. 
The formed in-homogeneous pattern (Fig.~\ref{ContinuumModelFig}ab)is comparable with that of the microscopic model. We found that the instability requires a large enough system size, and the critical size increase with the response time scale (Fig.~\ref{ContinuumModelFig}c). In the large $\tau$ limit, the critical system size $L_c$ appears to be proportional to $\sqrt{\tau}$. 
 
\section{\label{sec:twostatemodel} The simplified two-state model}
The model proposed in the previous section was still difficult to tackle analytically. In order to have deeper understanding of the role of the response time, we propose a simplified two-state model that capture the important features, and perform the stability analysis.

Let us consider a run-and-tumble model where a cell can take either the high tumbling state with the tumbling frequency $\alpha_h$ or the low tumbling state with the tumbling frequency $\alpha_l$. In each state, a cell will "run" at a velocity $v$ between the tumbling, and switch the direction with the given tumbling frequency. We introduce the density-controlled average tumbling frequency by assuming that the ratio $r(\rho)$ between the low frequency state and the high frequency state in the steady state is given by a function
\begin{equation}
r(\rho)=\frac{1}{1+(\rho/\rho_c)^n},\label{rrho}
\end{equation}
with $\rho$ represents the local density of the bacteria, making the average tumbling frequency $\bar \alpha(\rho)$ for a given density $\rho$ to be 
\begin{equation}
\bar \alpha(\rho)=r(\rho)\alpha_l+(1-r(\rho))\alpha_h.
\end{equation}
To introduce the time scale of the state change, $\tau$, we introduce the rate for an bacterium to switch from the high (low) tumbling state to the low (high) tumbling state to be $r(\rho)/\tau$ [$(1-r(\rho))/\tau$].

By introducing the density of bacteria in the low (high) state $\rho_l$ ($\rho_h$), we get
\begin{eqnarray}
\rho&=&\rho_h+\rho_l, \label{2state1}\\
\frac{\partial \rho_l}{\partial t }&=&-\frac{\partial J_l}{\partial x}-\frac{1}{\tau}\left[\left(1-r(\rho)\right)\rho_l-r(\rho)\rho_h\right],\label{2state2}\\
\frac{\partial J_l}{\partial t }&=&-v^2\frac{\partial \rho_l}{\partial x}-\frac{1}{\tau}\left[\left(1-r(\rho)\right)J_l-r(\rho)J_h\right]\nonumber \\
&& -2\alpha_l J_l,\label{2state3}\\
\frac{\partial \rho_h}{\partial t }&=&-\frac{\partial J_h}{\partial x}+\frac{1}{\tau}\left[\left(1-r(\rho)\right)\rho_l-r(\rho)\rho_h\right],\label{2state4}\\
\frac{\partial J_h}{\partial t }&=&-v^2\frac{\partial \rho_h}{\partial x}+\frac{1}{\tau}\left[\left(1-r(\rho)\right)J_l-r(\rho)J_h\right]\nonumber \\
&& -2\alpha_h J_h.\label{2state5}
\end{eqnarray}

Note that, in the limit of infinitely fast response ($\tau=0$) where
$\rho_l=r(\rho)\rho$, $\rho_h=\left(1-r(\rho)\right)\rho$, 
$J_l=r(\rho)J$, 
and $J_h=\left(1-r(\rho)\right)J$
always hold, eqs.~(\ref{2state1})-(\ref{2state5}) reduces to the no time delay case of eqs.~(\ref{eq:rho}) and (\ref{eq:J}) where $\alpha(h)$ replaced with $\bar \alpha(\rho)$.
Clearly the model does not support the non-uniform steady state without the time delay. 

The two-state model (\ref{2state1})-(\ref{2state5}) has a uniform steady state solution 
\begin{eqnarray}
\rho_l&=&r(\rho_0)\rho_0, \label{uni1}\\
\rho_h&=&\left(1-r(\rho_0)\right)\rho_0,\\
J_h&=&J_l=0.\label{uni4}
\end{eqnarray}

We now perform the linear stability analysis of the uniform steady state solution (\ref{uni1})-(\ref{uni4}).
By setting $r_0=r(\rho_0)$, we assume $\rho_l=r_0\rho_0+\tilde \rho_l$, 
$\rho_h=\left(1-r_0\right)\rho_0+\tilde \rho_h$,
$J_h=\tilde J_h$, $J_l=\tilde J_l$, linearize over the perturbation,
and look for the normal modes in the form
\begin{equation}
\left(\begin{array}{c}
\tilde \rho_l\\
\tilde J_l\\
\tilde \rho_h\\
\tilde J_h
\end{array}\right)=
\left(\begin{array}{c}
\hat \rho_l\\
\hat J_l\\
\hat \rho_h\\
\hat J_h
\end{array}\right)\exp\left(\omega t-ikx\right).
\end{equation}

\begin{figure}
\includegraphics[width=0.4\textwidth]{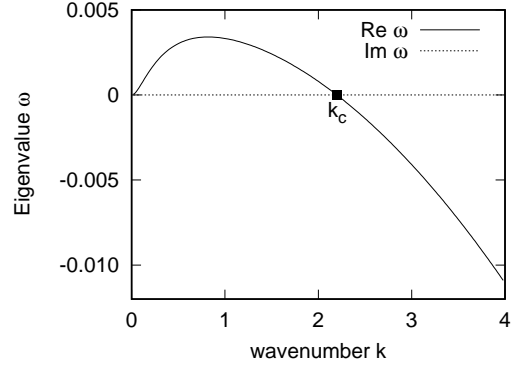}
\caption{The eigenvalue $\omega$ for the most unstable mode as a function of the wave number $k$.
The response time $\tau$ is set to 400s, and the Hill coefficient $n$ is set to $10$. The mean density  $\rho_0$ is set to be $\rho_c$. Solid line shows the real part, which increase in proportional to $k^2$ in the small-$k$ limit. The critical wave number $k_c$ is given by eq.~(\ref{kc}).
Dotted line shows the imaginary part of the mode, which stays zero.
}
\label{dispersion}
\end{figure}
\begin{figure}
\includegraphics[width=0.4\textwidth]{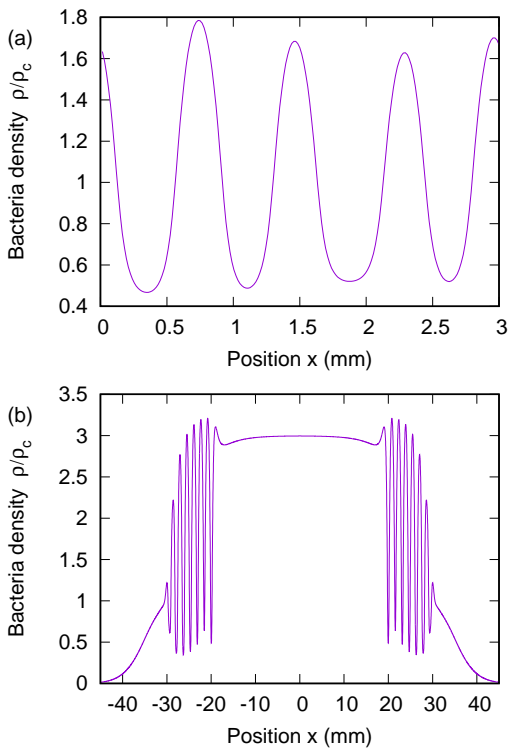}
\caption{(a) Numerical simulation of two state model with $\alpha_h=11.25s^{-1}$,$\alpha_l=0.25s^{-1}$, $v=15\mu ms^{-1}$, $\tau=4\times 10^{2}s$. Initial condition was $\rho_{l}=r(\rho_0)\rho_0$, $\rho_{h}=(1-r(\rho_0))\rho_0$, $J_l=J_h=0$.
Small randomness was added to the initial density by adding uniformly distributed number in the range $(0, 2.5\times 10^{-4}\rho_0)$ to $\rho_l$ and $\rho_f$.
(b)Two state model with growth. $\alpha_h=11.25s^{-1}$, $\alpha_l=0.25s^{-1}$, $v=15\mu ms^{-1}$, $\tau=15$min, 
and $\rho_{max}=3\rho_c$ are used. The growth rate $g$ is set to $1/\tau$. Non-linearity of the response $n$ is set to 10. 
For the initial condition, a small population of bacteria was placed in the center at $t=0$ as follows:
$\rho_0(x)=\frac{d}{\sqrt{2\pi}}e^{-\frac{x^2}{2w^2}}$ with $d=0.1 \rho_c$, $w=0.96$mm, $\rho_{l}(x)=r(\rho_0(x))\rho_0(x)$,  $\rho_{h}(x)=(1-r(\rho_0(x))\rho_0(x)$, $J_h(x)=J_l(x)=0$.
The snapshot was taken at $t=1 \times 10^5$s$\sim$28 hours. 
Note that the stripe pattern disappears in the middle due to the logistic term that allows growth for low density region, as discussed in ref.~\cite{fu2012stripe}.}
\label{twostateFig}
\end{figure}
The dispersion relations can be obtained by using Mathematica, as an example of the most unstable mode is shown in Fig.~\ref{dispersion}.
The imaginary part of $\omega$ for the most unstable mode is zero, consistent with the fact that the pattern does not travel. 
Fig.~\ref{dispersion} clearly shows that the instability happens at the long-wavelength (small $k$). 

By expanding the dispersion relation for the small $k$, we get an analytical expression
\begin{equation}
\omega(k)=-\frac{1+2\alpha_l \tau+2(\alpha_h-\alpha_l)\tau(r_0+r'_0\rho_0)}{2\alpha_l r_0+2\alpha_h(1-r_0)+4\alpha_l\alpha_h\tau}v^2k^2+O(k^3),
\label{omegakToSecond}
\end{equation}
where $r'_0=r'(\rho_0)$ with prime being the derivative by its argument. 
The sign of the $k^2$ term determines the stability in the long wave length. Note that when $\tau=0$, $\omega(k)=-\frac{v^2}{2\bar \alpha(\rho_0)}k^2+O(k^3)$, i.e., the system is always linearly stable in the long wavelength limit and the stability is characterized by the diffusion constant for the given density $v^2/2\bar \alpha(\rho_0)$. 
When $\tau>0$ and $r'_0<0$, the sign can be positive, i.e., the system has effectively a negative diffusion constant that makes the uniform state unstable. 

Especially, at $\rho_0=\rho_c$, the expression simplifies, and the condition for the instability can be written as  
\begin{equation}
n>\frac{2}{(\alpha_h-\alpha_l)\tau}+\frac{4}{\alpha_h/\alpha_l-1}+2.
\end{equation}
This shows that the non-linearity of $\rho$ dependence needed for the instability 
diverges if we take $\tau\to +0$ limit with keeping the tumbling frequency difference $(\alpha_h-\alpha_l)$ finite, consistent with the lack of instability without the delay.
At the same time, with realistic parameters for {\it E. coli}, $\alpha_h-\alpha_l$ is order of 10/sec and $\alpha_h/\alpha_l$ is order 10, so as long as the delay $\tau$ is order of a few seconds or larger, the hill coefficient $n$ slightly larger than 2 will be sufficient for instability.

The critical wave-number $k_c>0$ above which the system stays stable can be calculated from $\omega(k_c)=0$, 
and given by 
\begin{equation}
k_c(\rho_0)=\frac{\sqrt{-\left[1+2\alpha_l\tau+2(\alpha_h-\alpha_l)\tau(r_0+r'_0\rho_0)\right]}}{\tau v}, 
\label{kc}
\end{equation}
which is a positive real number when the system is linearly unstable. 
Eq.~(\ref{kc}) shows that the smallest system size $L_c=2\pi/k_c$ necessary to see the instability increases with $\tau$, and scales with $\sqrt{\tau}$ in the large $\tau$ limit. This is consistent with the observation in the continuum model as shown in Fig.~\ref{ContinuumModelFig}c.

At $\rho_0=\rho_c$, we have
\begin{eqnarray}
L_c(\rho_c)&=&\frac{2\pi \tau v}{\sqrt{-\left[1+2\alpha_l \tau+(\alpha_h-\alpha_l)\tau(2-n)/2\right]}},\nonumber\\
\label{lcrhoc}
\end{eqnarray}
showing that the critical system size also dependent on the non-linearity of the response $n$. 

We confirmed that the two state model simulated with the parameters where the uniform solution is linearly unstable
shows a stripe pattern as shown in Fig.~\ref{twostateFig}a. 
We observe the formation of the high and low density regions, similar to the pattern observed in the original complex model (Fig.~\ref{ContinuumModelFig}).

Finally, we checked if the two-state model can give a stripe pattern in growing bacterial front, when the time delay is comparable with the bacterial division time, which would be the situation where the time delay is coming from the dilution of the stable proteins. 
For this purpose, we simulate two-state model with the growth  
by adding a logistic growth term $g\rho_l(1-\rho/\rho_{max})$ to eq.~(\ref{2state2})
and $g\rho_h(1-\rho/\rho_{max})$ to eq.~(\ref{2state4}), respectively,
with a growth rate $g$ (cf.~\cite{fu2012stripe}). We set $g=1/\tau$ and simulated the expansion of the bacterial population in one dimension. 
The result shown in Fig.~\ref{twostateFig}(b) confirmed that the system can obtain the stripe formation at the front with a high enough non-linearity. 

\section{Summary and Discussion}
We have shown that adding a response time delay to a simple run-and-tumble model with bacterial density dependent tumbling frequency causes the instability in the uniform density profile. The instability is long-wave length, and the critical system size $L_c$ for instability grows with the time delay as $L_c\propto \sqrt{\tau}$ in the large $\tau$ limit. These observations are reproduced in analytically tractable two state model, where the explicit parameter dependence of $L_c$ was obtained. For the two state model, it was found that the non-linearity $n$ required for the instability diverges in the $\tau\to +0$ limit. This is consistent with the previous finding of lack of instability in  run-and-tumble model with $\tau=0$. Finally, the two state model was extended to include the cellular growth, and it was confirmed that the stripe formation is possible even if the time delay $\tau$ is comparable with the generation time $1/g$. 
The stripe formation was seen in the growth front, showing that instability is there for this relatively long time delay. 

In biological systems, the typical timescales for different types of regulations can vary significantly, and sometimes significant delay in response can happen. The present work demonstrates the importance of taking into account those delays in proper modeling of observed phenomena.

\section*{Acknowledgement}
This work was funded by the Danish National Research Foundation (BASP: DNRF120). XZ and NM thank Leihan Tang for fruitful discussions. 
\bibliographystyle{apsrev4-1}
\bibliography{stripe}
\end{document}